\documentclass[aps,pra,twocolumn]{revtex4-1}
\usepackage{graphicx}
\usepackage{epsfig}
\usepackage{dcolumn}
\usepackage{bm}
\usepackage{color}

\usepackage{float}
\usepackage{amsmath}

\begin{document}

\title{Conditional emergence of classical domain and branching of quantum histories}

\author{Alexei V. Tkachenko}
\affiliation{  Center for Functional Nanomaterials, 
Brookhaven National Laboratory, Upton NY 11973}
\email{oleksiyt@bnl.gov}

\begin{abstract}

We outline the Minimalistic Measurement Scheme (MMS) compatible with regular unitary evolution of a closed quantum system.  Within this approach, a part of the system becomes informationally isolated (restricted) which leads to a natural emergence of  the classical domain. This measurement scenario is a simpler alternative to environment-induced decoherence. In its basic version, MMS involves two ancilla qubits,  $A$ and $X$, entangled with each other and with the System $S$. Informational or thermodynamic cost of measurement is represented by $X$-qubit being isolated, i.e. becoming unavailable for future interactions with the rest of the system.  Conditional upon this isolation, $A$-qubit, that  plays the role of an Apparatus, becomes classical and records the outcome of the measurement. The procedure may be used to perform von Neumann-style projective measurements or generalized ones, that  corresponds to  Positive-Operator Value Measure (POVM).  By repeating the same generalized measurement multiple times with different $A$- and $X$-qubits, one asymptotically approaches the  wave function collapse in the basis determined by the premeasurement process. We present a simple result for the total information extracted after $N$ such {\it weak} measurements. Building upon  MMS, we propose a construction that maps a  history of a quantum system onto a set of $A$-qubits. It resembles the Consistent History (CH) formulation of Quantum Mechanics (QM), but is distinct from it, and is built entirely within the conventional QM. In particular, consistency postulate of CH formalism is not automatically satisfied, but rather is an emerging property. Namely, each measurement event corresponds to the branching of  mutually exclusive  classical realities whose probabilities are additive.  In a general case, however,  the superposition between different histories is determined by the {\it history density matrix}.      
\end{abstract}
\maketitle

\section{Introduction}

As it approaches its centennial, Quantum Mechanics (QM) is still commonly perceived as a counterintuitive and mysterious field of Physics. Most of its postulates and the overall mathematical structure,   though exotic at the time of their development, are relatively straightforward.  However, there is one important element of QM  that remains puzzling and controversial, and is responsible for many of its paradoxical aspects: the measurement problem \cite{vonNeu,Dirac:1930:PQM,Bohm,Everett, Wheeler, Wigner}. In QM, the state of the system is describe as a vector in  Hilbert space, called wave function $| \Psi \rangle$, and its  time evolution is given by a  linear unitary operator: $| \Psi \rangle \rightarrow \widehat T| \Psi \rangle $. This evolution can be calculated, e.g. with the help of  Schr{\"o}dinger Equation, Heisenberg's Matrix Mechanics or Dirac-Feynman path integrals.
According to the conventional Copenhagen formulation of  QM, at the moment of measurement,  the regular unitary evolution stops working and the wave function "collapses", i.e. the system ends up in one of the eigenstates of the corresponding operator.   This prescription is practical, but conceptually problematic since the observer is assumed to live in a classical world rather than being described by QM itself.  

There is a long history of research into the topic of quantum measurement, starting with  John von Neumann's scheme proposed in the early days of QM \cite{vonNeu}. He demonstrated how a measuring apparatus operating according to the laws of QM, can be inserted between the measured system and the observer.  He also formalized the wave function collapse process in the form of the {\it Projection Postulate}. According to it, the wave function, at the time of measurement, changes as $|\Phi\rangle \rightarrow \widehat{P}_k|\Phi\rangle/\sqrt{p_k}$  where    $\widehat{P}_k=|k\rangle \langle k|$ is the projection operator associated with the observed  eigenstate $| k\rangle$, and $p_k= \langle  \Phi|\widehat{P}_k|\Phi\rangle$ is the probability of that particular outcome. 
Von Neumann's theory does not resolve the quantum measurement problem but rather restates mathematically the Copenhagen-style prescription of wave function collapse.  In particular,  von Neumann argues that an interface between the quantum and classical worlds is ultimately unavoidable since any measurement has to be eventually perceived by an observer who lives in a classical reality. Attempts to resolve the quantum-classical conundrum included an exotic attempt by  Bohm to couple wave function to real particle dynamics \cite{Bohm}, as well as an esoteric multi-world interpretation by Everett \cite{Everett,Wheeler}, in which observer's own brain  becomes a part of the theory.  

An important breakthrough came with  the so-called decoherence program  introduced in the works of Zurek, Gell-Mann, Hartle,  and others \cite{Zurek_RMP, Zurek_PT,Hartle,Schloss_decoh}. It is based on the observation that any practical measurement device interacts with an environment. The decoherence associated with this interaction explains, at least partially, the branching of the system between different classical realities that represent different outcomes of the measurement.  However, the perfect environment-induced measurement is only achieved in the thermodynamic limit. So,  even the simplest quantum system could be described self-consistently only  as a part of an infinitely large one.  As a result, within the decoherence program,  the  fundamental problem of quantum measurement becomes dependent on our ability to understand statistical properties of a complex system. Partially because of this, its generality and limits of applicability are not fully established. 

In this paper, we propose a simple, self-consistent description of quantum reality which is built strictly within the conventional QM, yet without the introduction of the wave function collapse or projection postulate. We also do not include any environment-induced decoherence, primarily for the sake of greater simplicity. Based on quantum information theory we argue that a measurement process unavoidably requires a loss of information. Within our approach, the  Minimalistic Measurement Scheme (MMS),   this informational sacrifice for each measurement is  represented by a single qubit that becomes informationally isolated, i.e. unavailable for any future interactions. Conditional upon its isolation, the classical domain naturally emerges within a quantum system. In Section III we describe MMS and demonstrate how it can be used to make  von Neumann-style projective measurements, as well as the generalized  quantum measurements, described as Positive-Operator Valued Measure (POVM). Building upon this approach, in Section IV we present a construction that maps a history of a quantum system onto a set of ancilla qubits that are consequently subjected to MMS. Our approach resembles but is not equivalent to Consistent Histories (CH) formalism. While CH is built upon a set of postulates of its own, our construction is done within the traditional framework of QM, with measurements implemented through MMS. In particular, one of the central elements of CH, consistency postulate, is not automatically satisfied. Instead, it is an emerging property, conditional upon informational isolation of a part of the system, as any classical information within MMS. Within the original CH approach, a  quantum history is defined as a chain of projections and unitary evolution operators. In our construction, we generalize this definition by replacing projectors with Kraus operators associated with POVM-like weak measurements.

\section{Von Neumann measurement theory and decoherence program}

To model  the measuring process,  von Neumann considered a combination of two quantum subsystems:  the System   $S$ , and the Apparatus $A$ \cite{vonNeu}. During the first step, which is called  premeasurement, a quantum entanglement between these two subsystems is achieved. Namely, if $S$ is in a quantum state   $\sum_k c_k | k \rangle$, and the Apparatus is originally in  state $|0\rangle_A $, the premeasurement is the following  unitary transformation:
\begin{equation}
\label{premeasure}
\left (\sum_k c_k | k \rangle\right) |0\rangle_A \rightarrow | \Psi \rangle= \sum_k c_k | k \rangle | k \rangle_A
\end{equation}
Here $| k \rangle $ and $| k \rangle_A$ are states of $S$ and $A$ respectively. Following the premeasurement, the wave function collapse is described as  a non-unitary transformation of the density operator, $\widehat{\rho} = | \Psi \rangle \langle  \Psi|$:  
\begin{equation}
\label{collapse}
\widehat{\rho} \rightarrow  \sum_k \widehat{P}_k\widehat{\rho}\widehat{P}_k
\end{equation}
Here $\widehat{P}_k=|k\rangle_A \langle k|_A$ are projection operators of the Apparatus subsystem. If both the premeasurement  Eq. (\ref{premeasure}), and the non-unitary projection process Eq. (\ref{collapse}), are performed in the same basis 
of the Apparatus states, $| k\rangle_A$, they would transform the density operator of the combined system   into the diagonal form:   
\begin{equation}
\label{diag}
\widehat{\rho} = \sum_k\left| c_k \right |^2  | k \rangle | k \rangle_A \langle  k|_A \langle  k|
\end{equation}
Here $\left| c_k \right |^2$ are  probabilities of different results, and since  all the off-diagonal  terms are zeros, there is no interference between those outcomes, i.e. the  superposition principle of the classical probability theory is recovered.

The non-unitary projection process is certainly inconsistent with conventional quantum dynamics. A reasonable justification for it was given much later,  within   the  decoherence program \cite{Zurek_RMP, Zurek_PT,Hartle,Schloss_decoh}.  Let us imagine that an environment  is coupled to the apparatus in such a way that the states  $| k \rangle_A$ of the latter are preserved in time, but the evolution of the environment would depend on  $| k \rangle_A$. As a result, following the premeasurement, Eq. (\ref{premeasure}) , the composite $S+A+E$ system will evolve after some time into a new state: 
 \begin{equation}
\label{envir}
| \Psi \rangle |0 \rangle_E=\left (\sum_k c_k | k \rangle | k \rangle_A \right ) |0\rangle_E \rightarrow  \sum_k c_k | k \rangle | k \rangle_A |k\rangle_E
\end{equation}

The density operator of $S+A$ system is obtained  by taking a partial trace of the overall $S+A+E$ density operator with respect to environment variables, 
$
\widehat{\rho}_{sa}={\bf Tr}_{e} \widehat{\rho}_{sae}
$  
At this stage, one would recover the von Neumann non-unitary measurement process, Eq. (\ref{collapse}), {\it if}  the eventual states of  environment  that correspond to different indexes  $k$ are mutually orthogonal: $\langle k |k'\rangle_E=\delta _{kk'}$.   This orthogonality condition  can indeed be proven directly for certain explicit  models of the environment.  

\section{Minimalistic Measurement Scheme}
\subsection{Conditional emergence of classical domain}
An important insight into the nature of a quantum measurement  is given by the quantum information theory \cite{vonNeu,CerfPRL,CerfPhysD,correlationRMP}. Quantum information entropy, also introduced by von Neumann,  $S=-{\bf Tr}   (\widehat{\rho} \log_2 \widehat{\rho}) $  is zero for the system in a  pure quantum state, and  is conserved by unitary evolution. Therefore, if the  initial quantum state of an isolated  system is known,  its information entropy is  $S=0$. If one now performs a new measurement on the same system  and records its result with a classical bit, the information entropy associated with that bit is $\Delta S=-p \log_2  p-(1-p)\log_2(1-p)$, where $p$ is the probability of it being in state $1$. After an observer reads the bit, this information is being extracted, and the measured system once again returns  to a pure state with $S=0$.  We come to a seemingly paradoxical conclusion that the new information is extracted from nowhere. The non-unitary von Neumann process and  the decoherence program both provide a partial resolution to this paradox.  In both cases, the information entropy of the system is being  {\it increased} during  the measurement. This allows one to reduce $S$ back to zero once the result of the measurement is read, and to record  the  new information about the system. In other words, either processes is needed  to {\it erase} some information. 
\begin{figure}[!t!h]
\includegraphics[width=0.5\textwidth]{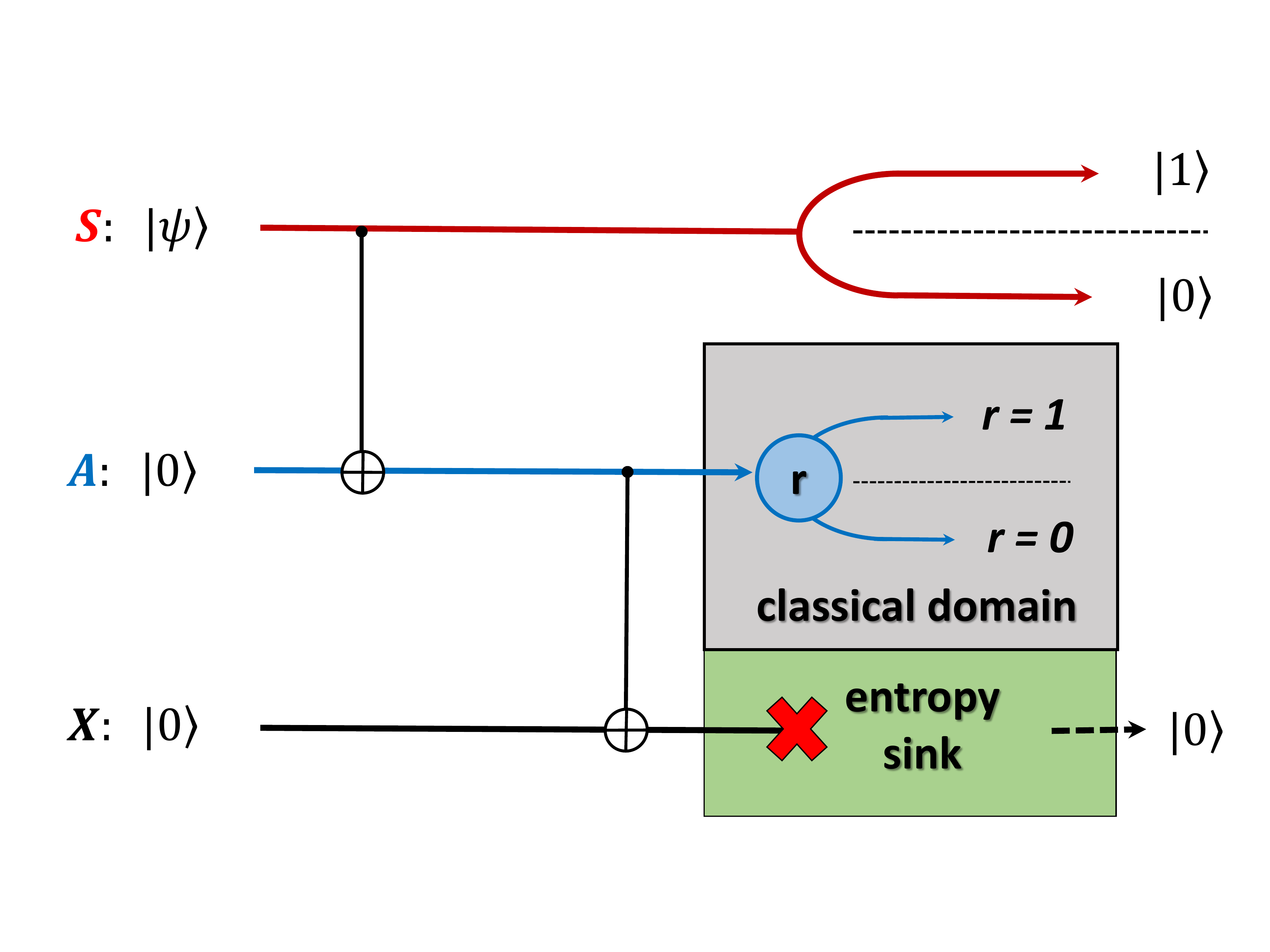}
    \caption{ Schematic representation  of Minimalistic Measurement Scheme (MMS)}
    \label{fig1:scheme}
\end{figure}

Here we introduce the Minimalistic Measurement Scheme (MMS) in which this sacrifice of information is  represented by a single qubit which becomes   unavailable for any future interactions. Consider a toy model  consisting of three qubits: System $S$, and two ancilla qubits , $A$ and $X$. The $A$-qubit plays the role of  an Apparatus and eventually will record the measurement outcome, while  $X$-qubit will be discarded.    Prior to the measurement, $S$ is in an unknown state $ |\psi\rangle= c_1 |1\rangle + c_0 |0\rangle$ while both   $A$  and $X$ are prepared in their respective "zero" states,  $|0\rangle_A$ and $|0\rangle_X$. The measurement protocol is as follows:  first, we perform the premeasurement with the help of quantum ${\bf cnot}$ gate  acting on   $S$ and  $A$, and then execute the same operation on   $A$ and $X$. ${\bf cnot}$ (or controlled-$NOT$) is one of the most common gates in quantum logic \cite{QComp}. When acting on the System and Apparatus with  our initial conditions,  it would results in von Neumann-style entanglement between them: $c_1 |1\rangle|1\rangle_A + c_0 |0\rangle|0\rangle_A$. Following this two-step process, all three qubits would become entangled:
\begin{equation}
\label{minimal1}
 |\psi\rangle |0\rangle_A |0\rangle_X    \rightarrow c_1 |1\rangle |1\rangle_A |1\rangle_X + c_0 |0\rangle |0\rangle_A |0\rangle_X  \end{equation}
This is a  unitary process   fully consistent with QM. Following it, we assume  $X$-cubit to become unavailable for any future interactions.       Mathematically, this would imply that instead of considering the full $S+A+X$ system, we should calculate the  density operator of the reduced $SA$ system by taking partial trace with respect to $X$-qubit: 
\begin{equation}
\label{rho}
\widehat{\rho}_{sa} = {\bf Tr}_X \widehat{\rho}_{sax}= \left| c_1 \right |^2  | 1 \rangle | 1 \rangle_A \langle  1|_A \langle  1|+ \left| c_0 \right |^2  | 0 \rangle | 0 \rangle_A \langle  0|_A \langle  0|
\end{equation}
This result is formally equivalent to the wave function collapse as given by  von Neumann's non-unitary projection process, Eq. (\ref{collapse}). The overall measurement process is schematically shown  in Fig. \ref{fig1:scheme}.

The classical domain within our  toy model   emerges as a direct consequence of ignoring information about $X$-qubit. There are several ways in which this process can be interpreted and/or implemented.   First,  one can employ an additional device that acts as the {\it entropy sink}. Its role is to supply  a new $X$-qubit in a pure quantum state $|0\rangle$, in exchange for the current one. Upon this exchange, we recover our original $S+A+B$ system, but the Apparatus now becomes a classical bit that has recorded the  result of the measurement, $r$, while  System $S$ ends up in a pure quantum state  $|r\rangle$ for each of the outcomes.  Information entropy associated with the measurement, $\Delta S= -\sum_r p_r\log_2 p_r$,  is equal to the one adsorbed by entropy sink, together with $X$-qubit.   If we consider $S+A+B$ as a single quantum system, its entropy would remain $0$. This means that the positive entropy $2\Delta S$ of the two separated qubits is equal and opposite to the entropy associated with the classical correlation and quantum entanglement between them (each contributing $-\Delta S$ to the overall quantum information  entropy) \cite{CerfPRL,CerfPhysD,correlationRMP,neg_entr,Thermo}. This  negative mutual entropy is  lost due to separation. Furthermore,  if  the design of the entropy sink is such that any information about $X$-qubit is getting effectively erased due to the underlying ergodicity \cite{thermal1,thermal2},  the ultimate entropy cost of a single-bit measurement is given by the maximum possible value of $\Delta S$, i.e. $1$. This result is consistent with  Landauer's Principle that sets the lower bound   for the  energy cost needed for  irreversible single-bit operation at finite temperature $T$: $E\min=k_BT\ln 2$ \cite{Landauer,Bennett_Land}. 

An alternative to the introduction of the entropy sink would be  the "gentleman's agreement", according to which $X$-qubit simply becomes unavailable for any future interactions.  Note that the choice between $A$ and $X$ is arbitrary: each of them can be used to record the result, while the other could be the "lost" qubit. Thus, one can imagine two independent observers, Alice and Bob, to have access to $A$- and $X$-qubits, respectively. This way, each of them would have a  classical record of the same measurement, as long as they are not allowed to communicate with each other. If the agreement is ever broken, i.e.  any information  is being exchanged between them, the non-classical correlations,  such  as violation of Bell's inequality, would become possible \cite{Bell1,Bell}.

The above procedure, while being relatively trivial, does provide a resolution to some of the issues related to the quantum measurement problem. In particular, the  combined $S+A+X$ system  follows a  regular unitary evolution at any time. The classicality of the Apparatus is an emerging property, subject to the condition of informational isolation of $X$-qubit after the measurement. The entanglement entropy between $A$ and $X$-qubits constitute the informational or thermodynamic  cost of the measurement. One aspect that this simplistic approach does not address is the so-called "superselection", i.e.  it does not differentiate between "observable" and "non-observable" quantum states.   Under the decoherence program, it is suggested that the basis of so-called "pointer" states  is chosen by means of   environment-induced superselection or  {\it einselection}. It implies that   the form of coupling between the Apparatus and the environment predetermines the observable which is being measured.

\subsection{Generalized measurements within MMS}
\begin{figure}[!t!h]
\includegraphics[width=0.5\textwidth]{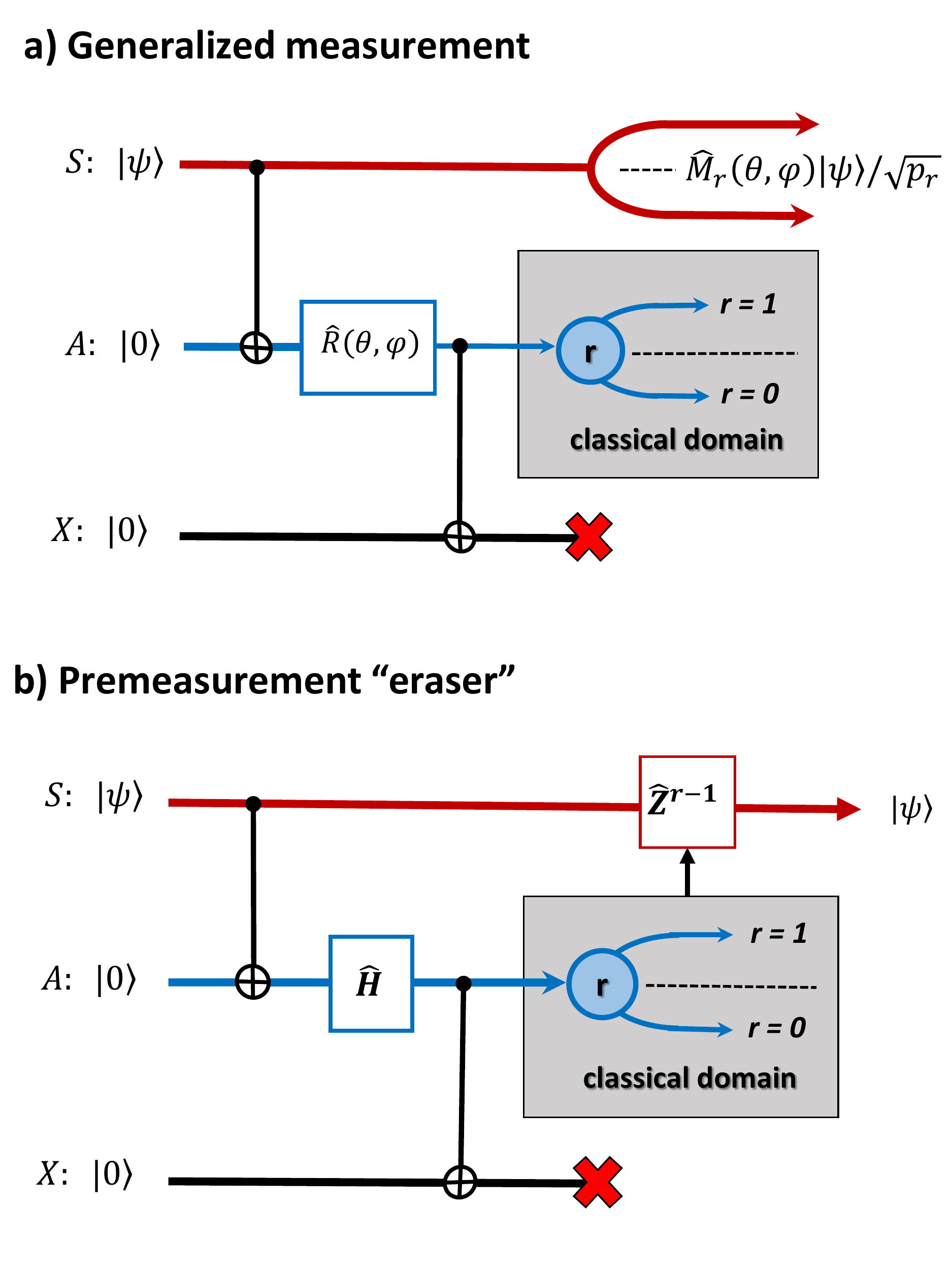}
    \caption{(a) Generalized quantum measurement with MMS.   (b) Erasing the premeasurement event.   $\widehat{H}=\widehat{R}(\pi/4,0)$ is Hadamard quantum gate. The outcome of the measurement  $r=1$ occurs with $50\%$ probability and indicates that the system has returned to its original state, with any effect of the premeasurement reverted. For $r=0$, the original state can be recovered with the help of Pauli-$Z$ gate.  }
    \label{fig2:erase}
\end{figure}

Now we consider a variation of the above  minimalistic scheme, in which the basis of the Apparatus states during the  two stages of the measurement process  are not the same. Specifically, following  the premeasurement, ${\bf cnot} (S,A)$, and prior to the  interaction with the $X$-qubit,  we  can apply a unitary rotation operator  to the Apparatus qubit: 
\begin{equation}
\label{rotation}
\widehat{R}_A(\theta,\phi)=\begin{bmatrix}
\cos \theta  & {\rm e}^{i\phi} \sin \theta \\
-{\rm e}^{-i\phi} \sin \theta & \cos \theta    
\end{bmatrix}_A
\end{equation}
After that,  $A$- and $X$-qubits are entangled by means of  ${\bf cnot}(A,X)$ quantum gate. As a result,  our original state is transformed   by three consecutive unitary operations, as shown in  Fig. \ref{fig2:erase}(a):  
\begin{equation}
\label{minimal2}
|\Psi\rangle_{\theta,\phi}= {\bf cnot} (A,X)  \times \widehat{R}_A (\theta,\phi) \times  {\bf cnot} (S,A) |\psi\rangle |0\rangle_A |0\rangle_X  
\end{equation}

As before, the  $X$-qubit is "discarded" and  we can calculate the resulting density operator of the $S+A$ system, by  taking partial trace over $X$ subsystem:
\begin{equation}
\label{rho1}
\widehat{\rho}_{sa} = {\bf Tr}_X |\Psi\rangle_{\theta,\phi}\langle\Psi|_{\theta,\phi} = \sum _{r=0,1} |r\rangle_A|S_r\rangle\ \langle S_r|\langle r|_A
\end{equation}
Here $|1\rangle_A$ and $|0\rangle_A$ form the basis  of the Apparatus states upon  rotation, Eq. (\ref{rotation}).  Just like in the previous case, the Apparatus  becomes  completely classical: there is no quantum interference between  $|1\rangle_A$ and $|0\rangle_A$.  This corresponds to the  two alternative  results of measurement, $r=1$ or $0$. The corresponding  System states are:
\begin{equation}
|S_r\rangle=\widehat{M}_r(\theta,\phi)|\psi\rangle
\end{equation}
Here $\widehat{M}_r(\theta,\phi)$ are known as Kraus operator \cite{Kraus} for outcomes $r=1,0$, respectively:
\begin{eqnarray}
\label{State}
\widehat{M}_r(\theta,\phi)= \sum_{k=1,0} R^*_{rk}(\theta,\phi) \widehat{P}_k= \nonumber\\
=\left\{\begin{array}{lr}
\widehat{P}_1\cos \theta   +  \widehat{P}_0{\rm e}^{-i\phi} \sin \theta   &  , r=1\\
- \widehat{P}_1{\rm e}^{i\phi} \sin \theta +\widehat{P}_0\cos \theta    &  , r=0 
\end{array}
\right.  \end{eqnarray}
 Here  $\widehat {R}^*(\theta,\phi)$  is  complex-conjugate to  rotation matrix $\widehat{R} (\theta,\phi)$ given by    Eq. (\ref{rotation}). 
The two state vectors,   $|S_1\rangle$ and $|S_0\rangle$, are in a general case not  orthogonal, and not yet normalized. The probabilities of  the corresponding outcomes can be found in a standard QM manner:  
\begin{equation}
p_r=\langle S_r|S_r\rangle 
\end{equation}

Importantly,  von Neumann's information entropy associated with the density operator $\widehat{\rho}_{sa}$ in Eq. (\ref{rho1}),  coincides with the Shannon's  entropy that can be calculated based on these probabilities:
\begin{equation}
S=-{\bf Tr} (\widehat{\rho}_{sa}\log_2 \widehat{\rho}_{sa})=-\sum_{r=0,1} p_r \log_2 p_r
\end{equation}
This means  that the density operator represents a so-called "ignorance interpretable" mixture of pure states. In other words, 
once  the result of the measurement is known, the System ends up in  pure quantum state, $|S_r\rangle / \sqrt{p_r}$.  However, in a general case,  the specific  states that correspond to each outcome  are not pre-determined by the measurement. Indeed, $\widehat{M}_r(\theta,\phi)$ in Eq. (\ref{State}) does not have the conventional form of a projection operator $\widehat P_k=|k\rangle \langle k |$,  so the final states will depend  on the state of the System prior to the measurement, $|\phi\rangle$. In particular,  one of the most important property of the projection operator, ${\widehat P}_k{\widehat P}_{k'}=\delta_{kk'}\widehat P_k $ is violated for $\widehat{M}_r(\theta,\phi)$.  This is an example of the generalized quantum measurement, associated with  Positive-Operator Valued Measure (POVM)  \cite{Kraus, NielsenChuang,Peres}. One of their signatures is that they are in general non-repeatable.  When two identical  measurements performed with  the same System but   with two different Apparatuses, they are not guaranteed to give the same result. This is a spectacular violation of the Projection Postulate, although the latter is still valid for the combined $S+A$ system. 

 An important difference from the regular  von Neumann projection  is that the  System, unlike the Apparatus,  is not  becoming classical immediately after the measurement, but  ends up  in a new coherent  quantum state. Furthermore, following the premeasurement process,  the $A$-qubit can in principle be stored under  decoherence-free conditions for as much time as feasible.  The measurement itself can be  done later, in any basis of one's choice.  One can even  completely "erase" the fact of the premeasurement \cite{eraser2000,eraser2002}.  In order to do so, the rotation operator must be given by  Hadamard quantum gate, $\widehat{H}=\widehat{R}(\pi/4,0)$. If now the result of the measurement is  $1$,  it would mean that any effect of the premeasurement on the System has been successfully reverted. This is because  $\widehat{P}_1(\pi/4,0)=\widehat{I}/\sqrt{2}$. The probability of that outcome is  $1/2$. Its alternative, $r=0$, also corresponds to a unitary transformation, Pauli-Z gate, applied to the System: $\widehat{P}_0(\pi/4,0)=(\widehat{P}_0-\widehat{P}_1)/\sqrt{2}=\widehat{Z}/\sqrt{2}$. If the result of the measurement is $0$, the System can still be  returned back to its original state by the same transformation, as shown in Fig. \ref{fig2:erase}(b).

\subsection{Asymptotic collapse of wave function}

In the above example, we returned the System back to its original state, and no information was extracted. On the other hand, the opposite limit when the measurement of the apparatus state is done in its original basis, as shown in Fig. \ref{fig1:scheme}, corresponds to traditional projection  measurement. The quantum state of the  System after the measurement is completely determined by the result, but its prior state is destroyed. A generalized measurement is shown in Fig. \ref{fig2:erase}(a) is an intermediate case. Only a fraction  of the information about the state of the system after the measurement is extracted,  and the system may still, in principle,   be returned to its original state $|\phi\rangle$. This is an example of a weak quantum measurement \cite{Weak_undoing2006,Weak_2008}. 

\begin{figure}[!t!h]
\includegraphics[width=0.5\textwidth]{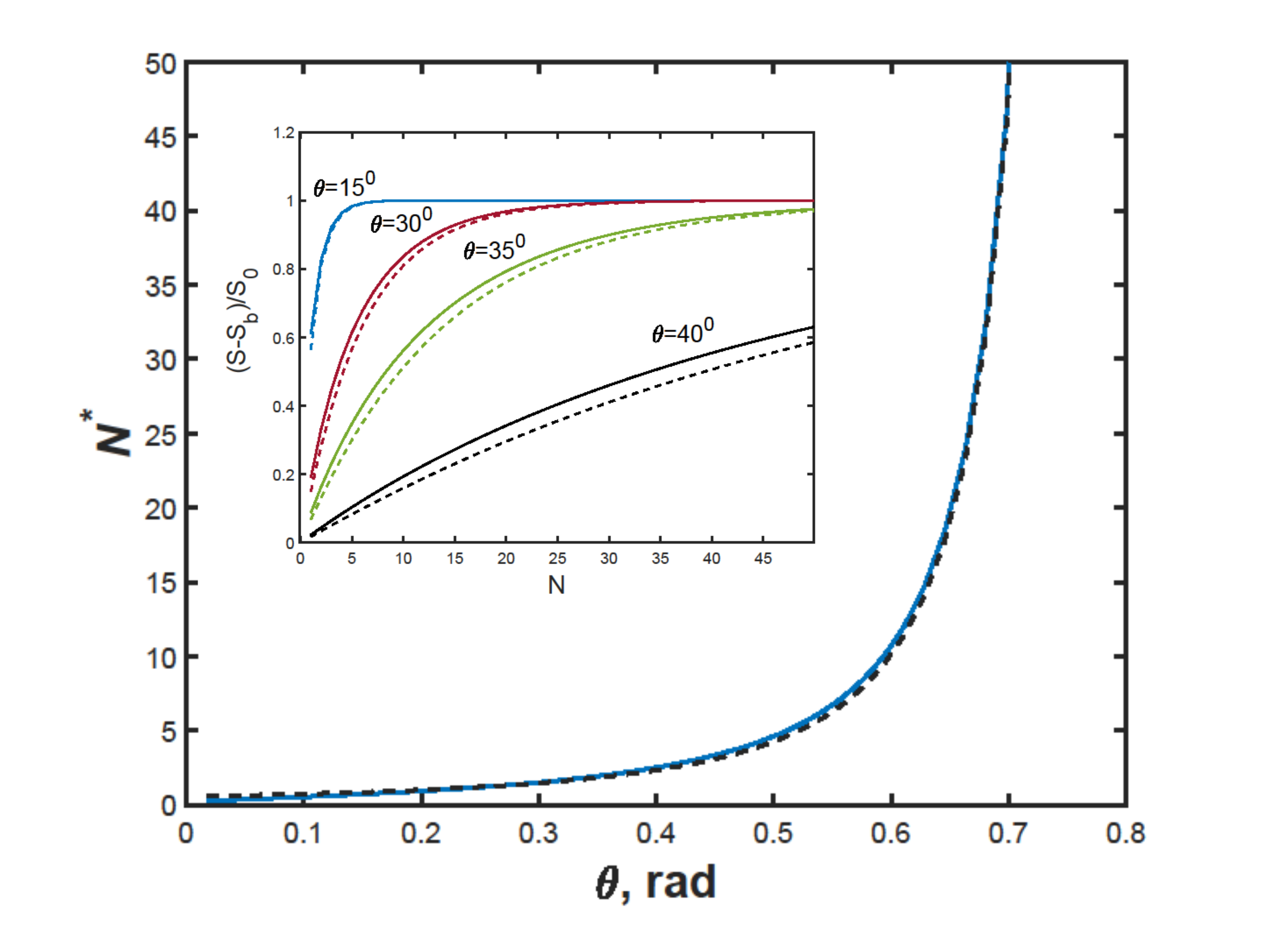}  
    \caption{Gradual "collapse" of the wave function  after multiple weak measurements. Characteristic number of measurements needed for information extraction, $N^*$ is plotted vs. rotation angle $\theta$ (solid line). Dashed line shows asymptotic result, Eq. (\ref{Nstar}). Insert: amount of information extracted vs. number of measurements $N$, for different values of $\theta$ (solid lines correwspond to  $|c_0|^2=|c_1|^2=1/2$, dashed - to  $|c_0|^2=0.1$).       }
    \label{fig3:collapse}
\end{figure}

By repeating the same weak  measurement for multiple times with different pairs of $A$- and $X$-qubits, one would achieve an asymptotic wave function collapse. It is in principle reversible \cite{Weak_undoing2006,Weak_2008}, though with exponentially diminishing probability.  Below we explicitly calculate the amount of  information extracted  after $N$ such  weak measurements are made. Note that  the number $l$ of   $A$-qubits that end up in state $r=1$ completely determine the overall Kraus operator associated with these $N$ measurements: 
\begin{equation}
\widehat{M}_l=\widehat{P}_1\cos^l \theta \left(-{\rm e}^{i\phi} \sin \theta\right)^{N-l}   +  \widehat{P}_0 \cos^{N-l} \theta\left({\rm e}^{-i\phi} \sin \theta\right)^l 
\end{equation}
The probability of a given value $l$ can be found as 
\begin{equation}
p_l={N \choose l}\left[|c_1|^2q^{2l}(1-q^2)^{N-l}+|c_0|^2q^{2(N-l)}(1-q^2)^{l}\right]
\end{equation}
Here $q=cos\theta$. One can see that this result  is a linear combination of two binomial distributions, $B_l(q^2,N)= {N \choose l} q^{2l}(1-q^2)^{N-l}$ and $B_l(1-q^2,N)$, with coefficients given by probabilities of the System to be in state $|1\rangle$ and $|0\rangle$, respectively. In the limit of large $N$ these distributions are strongly peaked at values $l_1=Nq^2$ and $l_0=N(1-q^2)$, which is an indication of an asymptotic wave function collapse in the basis determined by the premeasurement. In this limit, the information entropy of the overall distribution, $S= -\sum_lp_l\log_2 p_l$,  is given by 
\begin{equation}
 S \rightarrow S_b(q^2,N)-\sum_{k=1,0}|c_k|^2\log_2\left(|c_k|^2\right)
\end{equation}
Here the first contribution is the entropy of the binomial distribution, $S_b(q^2,N)=-\sum_l B_l(q^2,N)\log_2 B_l(q^2,N)$, and the second one is the total information entropy extracted from the System, $S_0$. In the insert of  Figure \ref{fig3:collapse} we show how much of this information is extracted, as a function of $N$, for various values of basis  rotation angle $\theta$.  One can see that quantity $(S-S_b(q^2,N))/S_0$ exponentially approaches $1$,  essentially independently of the original state of the System, i.e.   
\begin{equation}
    S-S_b(q^2,N)=S_0(1-\exp(-N/N^*)
\end{equation} 

Naturally, parameter $N^*$ diverges in the vicinity of $\theta=\pi/4$  since no information can be extracted at that  angle. One can determine the asymptotic behavior of $N^*$ for small values of $x=\theta-\pi/4\approx 1/2-q^2$. In order to do this, we consider the case when the two states of the system have the same probability (i.e. $S_0=1$), and  calculate the leading contribution in $x$ to the amount of entropy extracted after $N$ measurements: $S-S_b(q^2,N)\approx 2x^2N/\ln 2$. This leads to the following result for the characteristic number of the required weak measurements:
\begin{equation}
\label{Nstar}
    N^*\approx \frac{\ln 2}{2\left(\theta-\pi/4\right)^2}
\end{equation} 
Remarkably, this asymptotic formula gives a near perfect  fit across the whole range of $\theta$, as shown in Figure \ref{fig3:collapse}.

\section{Branching of quantum histories}

Now we consider a more general case: a System consisting of $d$ qubits  undergoes a unitary evolution. At certain moments of time  it gets  sequentially entangled  with $n$ different qubits: $A_1$,$A_2$,....,$A_n$, representing $n$ distinct premeasurement events.  In a general case,  these events are time ordered, but some of them may occur at the same time, or order within certain  sub-group of them may be not important. As discussed above, the $A$-qubit can in principle be stored for some time before the actual measurement is done. The resulting state of the System with all $n$  such qubits can be expressed as 
\begin{equation}
\label{history}
|\Psi \rangle= \sum_\alpha | \alpha_1\rangle_1 | \alpha_2 \rangle_2... | \alpha_n\rangle_n \widehat{C}_{\alpha}|\psi_0\rangle
\end{equation}
Here $| \alpha_l\rangle_l$ represents the sate of ancilla qubit $A_l$  ($\alpha_l=0 {\rm \ or \ } 1$),  and  $\alpha=\left(\alpha_1,\alpha_2,...,\alpha_n\right)$ is a vector  that represents the combined state of all $n$ $A$-qubits. $\widehat{C}_{\alpha}$ is known as a chain operator which is a series of projectors ${\widehat P}^{(l)}_{\alpha_l}$ and time evolution operators ${\widehat T}(t_{l-1}\rightarrow t_l)$  :
\begin{equation}
\label{chain}
\widehat{C}_{\alpha}=\prod_{l=1}^n{\widehat P}^{(l)}_{\alpha_l} {\widehat T}(t_{l-1}\rightarrow t_l)
\end{equation}

One may notice that our construction, Eqs.(\ref{history})-(\ref{chain}), strongly resembles the Consistent Histories (CH)  approach to QM \cite{hist_grif,hist_Isham,hist_omnes,decoh_hist,Hartle}. In its language, each vector $\alpha$ would represent a specific history of the quantum system. Despite this similarity,  there are important conceptual differences between our approach and CH formalism. First, our construction is done entirely within the framework of conventional QM (though without explicit  wave function collapse or projective postulate), while CH is built on its own set of postulates. Second, the fundamental space within CH is a tensor product of Hilbert Spaces of the original System at different times. In contrast, we simply expanded the System's Hilbert space at a given time by combining it with that of $n$  $A$-qubits. 

The overall density operator of the combined system,  ${\widehat \rho}=| \Psi \rangle\langle\Psi |=|\alpha\rangle \widehat{C}_{\alpha}{\widehat \rho}_0 \widehat{C}_{\alpha'}^*\langle \alpha'|$ can be expressed as a $2^n \times 2^n$ matrix in which each element is itself an  operator in the  System's Hilbert space: ${\widehat \rho}_{\alpha\alpha'}=\langle \alpha| {\widehat \rho}|\alpha'\rangle=\widehat{C}_{\alpha}{\widehat \rho}_0 \widehat{C}^*_{\alpha'}$. By taking a partial trace over System's final states, we obtain the reduced density operator that can be called {\it history density matrix}:
\begin{equation}
\label{densematr}
    D_{\alpha\alpha'}={\bf Tr}_S \left(\widehat{C}_{\alpha}{\widehat \rho}_0 \widehat{C}^*_{\alpha'}\right)
\end{equation}
Furthermore, our assumption that the system was prepared in a particular pure state $|\psi_0\rangle$, is in fact  redundant: if needed, the preparation can be implemented as a set of initial measurements. Under the assumption that  only $A$-qubits are available for measurements,  we should  take a partial trace over all plausible initial states of the system, which amount to setting its   a-priory density operator to ${\widehat \rho}_0=2^{-d}{\widehat I}$.  One of the central postulates of CH, known as  consistency condition, would require mutual orthogonality of different histories, i.e.  $D_{\alpha\alpha'}=0$  for $\alpha\ne \alpha'$. However, this condition is  not automatically  satisfied within  our construction, reflecting the constraints of the standard QM to which we adhere.  The history density matrix, Eq. (\ref{densematr}) is in fact a direct analog of Decoherence Functional used by Hartle and Gell-Mann for describing effect of coarse graning on quantum  histories in continuous space-time \cite{Hartle,decoh_hist}.  

\begin{figure}[!t!h]
\includegraphics[width=0.5\textwidth]{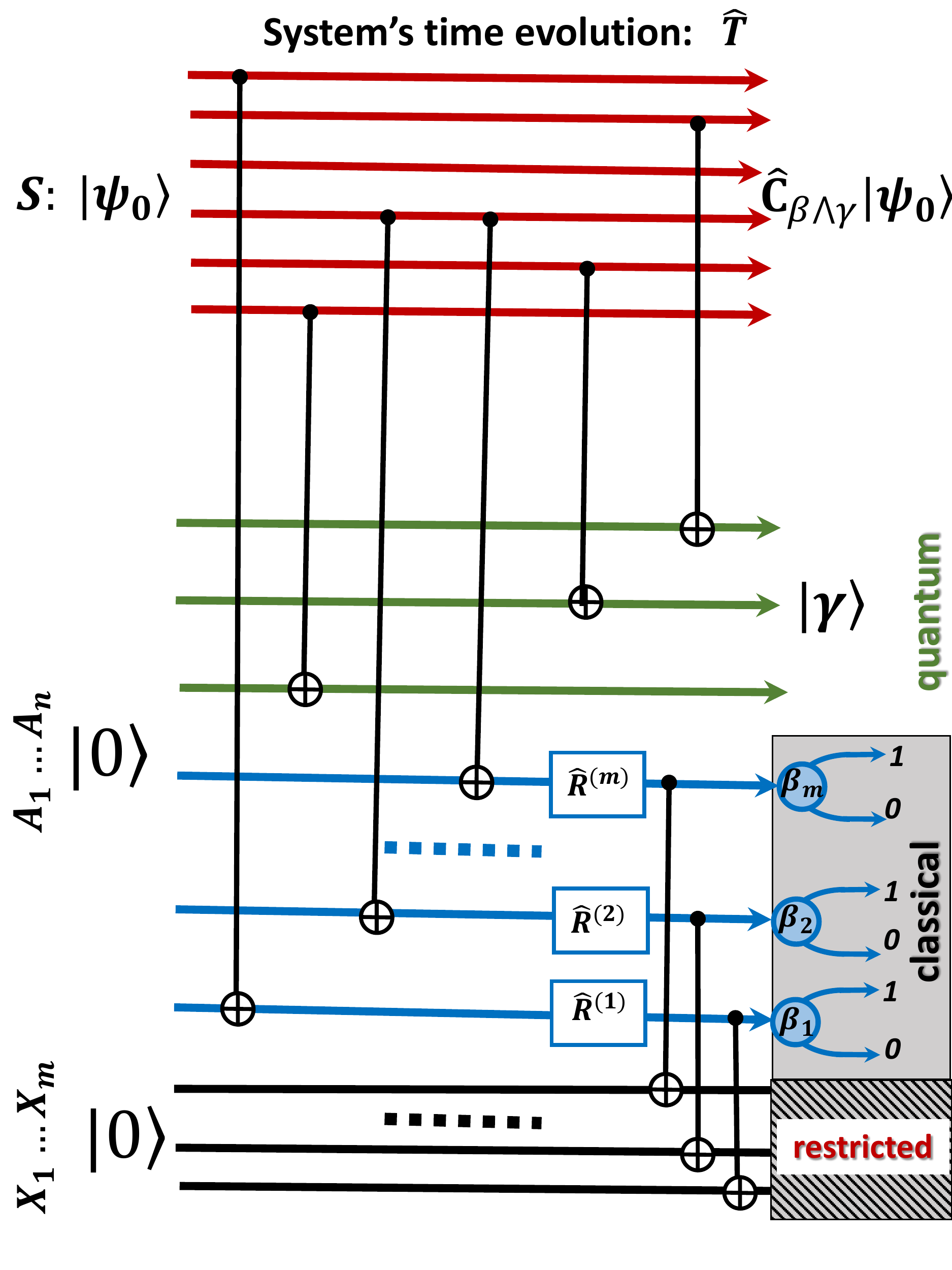}
    \caption{Schematic representation of quantum history construction, with MMS-based branching  of classical realities.}
    \label{fig4:history}
\end{figure}

We can now employ  MMS  described in the previous section: each $A$-qubit is first subjected to unitary transformation $\widehat R^{(l)}$, followed by entanglement with the respective $X$-qubit. If all $n$ of those become informationally isolated or discarded, the whole history vector  $\beta=(\beta_1,...,\beta_n)$ is measured. Following this, all the  off-diagonal elements of the new history density matrix $\widehat \mathbf{D}$ will be set to zero:   $\mathbf{D}_{\beta\beta'}= \delta_{\beta \beta'}{\bf Tr}( \widehat{\mathbf{C}}_{\beta}{\widehat \rho}_0 \widehat{\mathbf{C}}_{\beta}^*)$, which coincides with  the consistency postulate of CH formalism.  Note that  the chain operator has been redefined, and will now be called the {\it  generalized history operator}:
\begin{equation}
\label{chain2}
\widehat{\mathbf{C}}_{\beta}=\prod_{l=1}^n{\widehat M}^{(l)}_{\beta_l} {\widehat T}(t_{l-1}\rightarrow t_l)
\end{equation}
 Compared to the traditional definition of $\widehat{C}_\alpha$, Eq. (\ref{chain}), it incorporates the possibility of weak/generalized measurements. As one can see,  the   projectors ${\widehat P}$ have been replaced with  Kraus operators ${\widehat M}$  which   correspond to POVM rather than regular  projective measurements.  As has been shown   in the previous section, these operators are  related to  the rotations of the corresponding $A$-cubits:  
\begin{equation} {\widehat M}^{(l)}_{\beta_l}=\sum_{k=0,1}{\widehat R}_{ \beta_l k}^{(l)*}{\widehat P}^{(l)}_{k} 
\end{equation}

If only a subset of $A$-qubits has been subjected to MMS, a history can be represented as a logical conjuction $\beta\wedge \gamma $ of $m$ measured and $n-m$ unmeasured binary variables:  $\beta=(\beta_1,...\beta_m)$ and $\gamma=(\alpha_m+1,...\alpha_n)$, respectively (without loss of generality, we can re-order $A$-qubits so that the first $m$ are measured).  The overall construction is illustrated in Figure \ref{fig4:history}. 
After the measurement, the system branches onto $2^m$ distinct classical realities parameterized by vector $\beta$. Post-measurement history density matrix will have the following form: 
\begin{equation}
\mathbf{D}_{(\beta\wedge \gamma) (\beta'\wedge \gamma')}=\delta_{\beta \beta'}\mathbf{D}^{(\beta)}_{\gamma \gamma'}=\delta_{\beta \beta'}\sum_{\alpha, \alpha'}{\widehat \mathbf{R}}_{\beta\alpha}D_{\alpha\alpha'}{\widehat \mathbf{R}}_{\alpha'\beta}^*
\end{equation}
 Here ${\widehat \mathbf{R}}$ is the unitary transformation  composed of all $A$-qubit rotations prior to the measurement:
 \begin{equation}
 {\widehat \mathbf{R}}_{\alpha\beta}=\bigotimes_{l=1}^m {\widehat R}_{ \alpha_l \beta_l}^{(l)*}
\end{equation}
 Only those sub-histories that  correspond to different $\beta$ are guaranteed to have $0$ off-diagonal matrix element between them, and hence no quantum interference. However, different sub-histories for a given $\beta$ preserve coherence and do not satisfy the consistency postulate of CH formalism. This makes our framework more general. Note also, that we have not made any assumption about the time sequence of premeasurement events that correspond to vectors $\beta$ and $\gamma$. In other words, our construction allows to  treat any information about the System's history in the same way, regardless of whether it corresponds to its past, its  future, or any intermediate time. The term "branching" that we use here  is often associated with Everett's  multi-world interpretation of QM\cite{Everett}. In our  case, it refers to the emergence of distinct  classical realities under the condition that $X$-qubits remain restricted (i.e.  informationally isolated). Importantly, the branching in a general case occurs not in  time, but rather as a function of the amount of information gained about the system.  
 
 For a given classical  history $\beta$, we can determine the conditional probability of each sub-history $\beta \wedge \gamma$ that refines it: $ p\left( \gamma|\beta\right)=\mathbf{D}^{(\beta)}_{\gamma \gamma}/\bf Tr \left(\mathbf{D}^{(\beta)}_{\gamma \gamma'}\right) $.
If the outcome  of a particular measurement $\gamma_l$ is not known, the probability of a given history is determined by taking the respective partial trace of the history density matrix, ${\bf Tr}_{\gamma_l} \left(\mathbf{D}^{(\beta)}_{\gamma \gamma'}\right)$ or equivalently, by taking a sum of probabilities of all the compatible histories. Importantly, ignoring the result of a  measurement, or not measuring a specific $A$-qubit, {\it is not equivalent} to not making that measurement on the System. As has been demonstrated  in the previous section,  the effect of any single premeasurement can be erased and quantum interference effects recovered. In order to do this, the rotation matrix for the corresponding $A$-qubit has to be given by the Hadamard transformation: ${\widehat R}^{(l)}={\widehat H}$, and  result $\beta_l=1$ has to  be selected.

\section{Summary}

In the present work, we revisited the long-standing quantum measurement problem and the issue of the quantum-classical divide. We worked  within the conventional framework of QM, but dropped  the projection postulate or, equivalently, the wave function collapse prescription. Instead, we formulated the Minimalistic Measurement Scheme (MMS) that assumes that a part of a quantum  system becomes restricted, i.e. unavailable for any future interactions. This naturally leads to the emergence of a classical domain, conditional upon the informational isolation of the restricted sub-system. Specifically, we used two groups on ancilla cubits: $A$-type, and $X$-type. The former represents measurement Apparatuses that eventually record classical information about the System, once the latter becomes restricted or lost. The entropy increase due to the informational isolation of $X$-qubits  represents informational or thermodynamic  cost of performing the measurement and extracting new information about the system. The probabilistic nature of QM is a direct consequence of this information loss during the measurement process.

In addition to conventional von Neumann-style projective measurements, MMS naturally describes the so-called generalized or weak quantum measurents (POVM-type). Repeating this measurement multiple times, with different pairs of $A$ and $X$ qubits,  results in a gradual wave function collapse of the system, which can even be reversed, but with a diminishing probability. The basis in which the collapse would eventually occur is determined by the pre-measurement process.    

Building upon MMS, we made the construction that maps a history of a quantum system onto a set of $A$-qubits. This construction resembles the Consistent History (CH) approach to quantum theory but is built entirely within the framework of conventional QM. In particularly, a key element of CH formalism, the consistency postulate, is not automatically satisfied. Rather, different histories brunch into different classical realities only upon the use of MMS. Our framework deviates from  CH in several ways. First, it is constructed within a conventional single-time Hilbert space; second, the chain operator that defines a given history may include non-projective generalized measurements, and finally, the quantum coherence is preserved for sub-histories within each classically distinct history, i.e. they may violate consistency postulate of CH.   When MMS is applied to the set of $A$-qubits, the branching of different  classical realities occurs. Each of these realities corresponds to a particular generalized history operator. Importantly, our construction is nearly time-agnostic: while time ordering of certain premeasurement events is important, the subset of measured $A$-qubits (and the choice of basis for each of them)  can be absolutely arbitrary.

{\bf Acknowledgement} This research was conducted in the Center for Functional Nanomaterials, which is a U.S. DOE Office of Science User Facility, at Brookhaven National Laboratory under Contract No. DE-SC0012704.

\bibliography{quantum}

\begin{thebibliography}{33}%
\makeatletter
\providecommand \@ifxundefined [1]{%
 \@ifx{#1\undefined}
}%
\providecommand \@ifnum [1]{%
 \ifnum #1\expandafter \@firstoftwo
 \else \expandafter \@secondoftwo
 \fi
}%
\providecommand \@ifx [1]{%
 \ifx #1\expandafter \@firstoftwo
 \else \expandafter \@secondoftwo
 \fi
}%
\providecommand \natexlab [1]{#1}%
\providecommand \enquote  [1]{``#1''}%
\providecommand \bibnamefont  [1]{#1}%
\providecommand \bibfnamefont [1]{#1}%
\providecommand \citenamefont [1]{#1}%
\providecommand \href@noop [0]{\@secondoftwo}%
\providecommand \href [0]{\begingroup \@sanitize@url \@href}%
\providecommand \@href[1]{\@@startlink{#1}\@@href}%
\providecommand \@@href[1]{\endgroup#1\@@endlink}%
\providecommand \@sanitize@url [0]{\catcode `\\12\catcode `\$12\catcode
  `\&12\catcode `\#12\catcode `\^12\catcode `\_12\catcode `\%12\relax}%
\providecommand \@@startlink[1]{}%
\providecommand \@@endlink[0]{}%
\providecommand \url  [0]{\begingroup\@sanitize@url \@url }%
\providecommand \@url [1]{\endgroup\@href {#1}{\urlprefix }}%
\providecommand \urlprefix  [0]{URL }%
\providecommand \Eprint [0]{\href }%
\providecommand \doibase [0]{http://dx.doi.org/}%
\providecommand \selectlanguage [0]{\@gobble}%
\providecommand \bibinfo  [0]{\@secondoftwo}%
\providecommand \bibfield  [0]{\@secondoftwo}%
\providecommand \translation [1]{[#1]}%
\providecommand \BibitemOpen [0]{}%
\providecommand \bibitemStop [0]{}%
\providecommand \bibitemNoStop [0]{.\EOS\space}%
\providecommand \EOS [0]{\spacefactor3000\relax}%
\providecommand \BibitemShut  [1]{\csname bibitem#1\endcsname}%
\let\auto@bib@innerbib\@empty
\bibitem [{\citenamefont {von Neumann}(1955)}]{vonNeu}%
  \BibitemOpen
  \bibfield  {author} {\bibinfo {author} {\bibfnamefont {J.}~\bibnamefont {von
  Neumann}},\ }\href@noop {} {\emph {\bibinfo {title} {Mathematical Foundation
  of Quantum Mechanics}}}\ (\bibinfo  {publisher} {Princeton University
  Press},\ \bibinfo {address} {Princeton, NJ, USA},\ \bibinfo {year}
  {1955})\BibitemShut {NoStop}%
\bibitem [{\citenamefont {Dirac}(1930)}]{Dirac:1930:PQM}%
  \BibitemOpen
  \bibfield  {author} {\bibinfo {author} {\bibfnamefont {P.~A.~M.}\
  \bibnamefont {Dirac}},\ }\href@noop {} {\emph {\bibinfo {title} {The
  Principles of Quantum Mechanics}}}\ (\bibinfo  {publisher} {Clarendon
  Press},\ \bibinfo {address} {Oxford, UK},\ \bibinfo {year}
  {1930})\BibitemShut {NoStop}%
\bibitem [{\citenamefont {Bohm}(1952)}]{Bohm}%
  \BibitemOpen
  \bibfield  {author} {\bibinfo {author} {\bibfnamefont {D.}~\bibnamefont
  {Bohm}},\ }\href {\doibase 10.1103/PhysRev.85.166} {\bibfield  {journal}
  {\bibinfo  {journal} {Physical Review}\ }\textbf {\bibinfo {volume} {85}},\
  \bibinfo {pages} {166} (\bibinfo {year} {1952})}\BibitemShut {NoStop}%
\bibitem [{\citenamefont {Everett}(1957)}]{Everett}%
  \BibitemOpen
  \bibfield  {author} {\bibinfo {author} {\bibfnamefont {H.}~\bibnamefont
  {Everett}},\ }\href {\doibase 10.1103/RevModPhys.29.454} {\bibfield
  {journal} {\bibinfo  {journal} {Reviews of Modern Physics}\ }\textbf
  {\bibinfo {volume} {29}},\ \bibinfo {pages} {454} (\bibinfo {year}
  {1957})}\BibitemShut {NoStop}%
\bibitem [{\citenamefont {Wheeler}(1957)}]{Wheeler}%
  \BibitemOpen
  \bibfield  {author} {\bibinfo {author} {\bibfnamefont {J.~A.}\ \bibnamefont
  {Wheeler}},\ }\href {\doibase 10.1103/RevModPhys.29.463} {\bibfield
  {journal} {\bibinfo  {journal} {Reviews of Modern Physics}\ }\textbf
  {\bibinfo {volume} {29}},\ \bibinfo {pages} {463} (\bibinfo {year}
  {1957})}\BibitemShut {NoStop}%
\bibitem [{\citenamefont {Wigner}(1963)}]{Wigner}%
  \BibitemOpen
  \bibfield  {author} {\bibinfo {author} {\bibfnamefont {E.~P.}\ \bibnamefont
  {Wigner}},\ }\href {\doibase 10.1119/1.1969254} {\bibfield  {journal}
  {\bibinfo  {journal} {American Journal of Physics}\ }\textbf {\bibinfo
  {volume} {31}},\ \bibinfo {pages} {6} (\bibinfo {year} {1963})}\BibitemShut
  {NoStop}%
\bibitem [{\citenamefont {Zurek}(2003)}]{Zurek_RMP}%
  \BibitemOpen
  \bibfield  {author} {\bibinfo {author} {\bibfnamefont {W.~H.}\ \bibnamefont
  {Zurek}},\ }\href {\doibase 10.1103/RevModPhys.75.715} {\bibfield  {journal}
  {\bibinfo  {journal} {Reviews of Modern Physics}\ }\textbf {\bibinfo {volume}
  {75}},\ \bibinfo {pages} {715} (\bibinfo {year} {2003})}\BibitemShut
  {NoStop}%
\bibitem [{\citenamefont {Zurek}(1991)}]{Zurek_PT}%
  \BibitemOpen
  \bibfield  {author} {\bibinfo {author} {\bibfnamefont {W.~H.}\ \bibnamefont
  {Zurek}},\ }\href {\doibase 10.1063/1.881293} {\bibfield  {journal} {\bibinfo
   {journal} {Physics Today}\ }\textbf {\bibinfo {volume} {44}},\ \bibinfo
  {pages} {36} (\bibinfo {year} {1991})}\BibitemShut {NoStop}%
\bibitem [{\citenamefont {Gell-Mann}\ and\ \citenamefont
  {Hartle}(1993)}]{Hartle}%
  \BibitemOpen
  \bibfield  {author} {\bibinfo {author} {\bibfnamefont {M.}~\bibnamefont
  {Gell-Mann}}\ and\ \bibinfo {author} {\bibfnamefont {J.~B.}\ \bibnamefont
  {Hartle}},\ }\href {\doibase 10.1103/PhysRevD.47.3345} {\bibfield  {journal}
  {\bibinfo  {journal} {Physical Review D}\ }\textbf {\bibinfo {volume} {47}},\
  \bibinfo {pages} {3345} (\bibinfo {year} {1993})}\BibitemShut {NoStop}%
\bibitem [{\citenamefont {Schlosshauer}(2004)}]{Schloss_decoh}%
  \BibitemOpen
  \bibfield  {author} {\bibinfo {author} {\bibfnamefont {M.}~\bibnamefont
  {Schlosshauer}},\ }\href {\doibase 10.1103/RevModPhys.76.1267} {\bibfield
  {journal} {\bibinfo  {journal} {Reviews of Modern Physics}\ }\textbf
  {\bibinfo {volume} {76}},\ \bibinfo {pages} {1267} (\bibinfo {year}
  {2004})}\BibitemShut {NoStop}%
\bibitem [{\citenamefont {Cerf}\ and\ \citenamefont {Adami}(1997)}]{CerfPRL}%
  \BibitemOpen
  \bibfield  {author} {\bibinfo {author} {\bibfnamefont {N.~J.}\ \bibnamefont
  {Cerf}}\ and\ \bibinfo {author} {\bibfnamefont {C.}~\bibnamefont {Adami}},\
  }\href {\doibase 10.1103/PhysRevLett.79.5194} {\bibfield  {journal} {\bibinfo
   {journal} {Physical Review Letters}\ }\textbf {\bibinfo {volume} {79}},\
  \bibinfo {pages} {5194} (\bibinfo {year} {1997})}\BibitemShut {NoStop}%
\bibitem [{\citenamefont {Cerf}\ and\ \citenamefont {Adami}(1998)}]{CerfPhysD}%
  \BibitemOpen
  \bibfield  {author} {\bibinfo {author} {\bibfnamefont {N.~J.}\ \bibnamefont
  {Cerf}}\ and\ \bibinfo {author} {\bibfnamefont {C.}~\bibnamefont {Adami}},\
  }\href@noop {} {\bibfield  {journal} {\bibinfo  {journal} {Physica
  D-Nonlinear Phenomena}\ }\textbf {\bibinfo {volume} {120}},\ \bibinfo {pages}
  {62} (\bibinfo {year} {1998})}\BibitemShut {NoStop}%
\bibitem [{\citenamefont {Modi}\ \emph {et~al.}(2012)\citenamefont {Modi},
  \citenamefont {Brodutch}, \citenamefont {Cable}, \citenamefont {Paterek},\
  and\ \citenamefont {Vedral}}]{correlationRMP}%
  \BibitemOpen
  \bibfield  {author} {\bibinfo {author} {\bibfnamefont {K.}~\bibnamefont
  {Modi}}, \bibinfo {author} {\bibfnamefont {A.}~\bibnamefont {Brodutch}},
  \bibinfo {author} {\bibfnamefont {H.}~\bibnamefont {Cable}}, \bibinfo
  {author} {\bibfnamefont {T.}~\bibnamefont {Paterek}}, \ and\ \bibinfo
  {author} {\bibfnamefont {V.}~\bibnamefont {Vedral}},\ }\href@noop {}
  {\bibfield  {journal} {\bibinfo  {journal} {Reviews of Modern Physics}\
  }\textbf {\bibinfo {volume} {84}},\ \bibinfo {pages} {1655} (\bibinfo {year}
  {2012})}\BibitemShut {NoStop}%
\bibitem [{\citenamefont {Bennett}\ and\ \citenamefont
  {DiVincenzo}(2000)}]{QComp}%
  \BibitemOpen
  \bibfield  {author} {\bibinfo {author} {\bibfnamefont {C.~H.}\ \bibnamefont
  {Bennett}}\ and\ \bibinfo {author} {\bibfnamefont {D.~P.}\ \bibnamefont
  {DiVincenzo}},\ }\href {\doibase 10.1038/35005001} {\bibfield  {journal}
  {\bibinfo  {journal} {Nature}\ }\textbf {\bibinfo {volume} {404}},\ \bibinfo
  {pages} {247} (\bibinfo {year} {2000})}\BibitemShut {NoStop}%
\bibitem [{\citenamefont {del Rio}\ \emph {et~al.}(2011)\citenamefont {del
  Rio}, \citenamefont {Aberg}, \citenamefont {Renner}, \citenamefont
  {Dahlsten},\ and\ \citenamefont {Vedral}}]{neg_entr}%
  \BibitemOpen
  \bibfield  {author} {\bibinfo {author} {\bibfnamefont {L.}~\bibnamefont {del
  Rio}}, \bibinfo {author} {\bibfnamefont {J.}~\bibnamefont {Aberg}}, \bibinfo
  {author} {\bibfnamefont {R.}~\bibnamefont {Renner}}, \bibinfo {author}
  {\bibfnamefont {O.}~\bibnamefont {Dahlsten}}, \ and\ \bibinfo {author}
  {\bibfnamefont {V.}~\bibnamefont {Vedral}},\ }\href@noop {} {\bibfield
  {journal} {\bibinfo  {journal} {Nature}\ }\textbf {\bibinfo {volume} {474}},\
  \bibinfo {pages} {61} (\bibinfo {year} {2011})}\BibitemShut {NoStop}%
\bibitem [{\citenamefont {Parrondo}\ \emph {et~al.}(2015)\citenamefont
  {Parrondo}, \citenamefont {Horowitz},\ and\ \citenamefont {Sagawa}}]{Thermo}%
  \BibitemOpen
  \bibfield  {author} {\bibinfo {author} {\bibfnamefont {J.~M.~R.}\
  \bibnamefont {Parrondo}}, \bibinfo {author} {\bibfnamefont {J.~M.}\
  \bibnamefont {Horowitz}}, \ and\ \bibinfo {author} {\bibfnamefont
  {T.}~\bibnamefont {Sagawa}},\ }\href {\doibase 10.1038/nphys3230} {\bibfield
  {journal} {\bibinfo  {journal} {Nature Physics}\ }\textbf {\bibinfo {volume}
  {11}},\ \bibinfo {pages} {131} (\bibinfo {year} {2015})}\BibitemShut
  {NoStop}%
\bibitem [{\citenamefont {Steinigeweg}\ \emph {et~al.}(2014)\citenamefont
  {Steinigeweg}, \citenamefont {Khodja}, \citenamefont {Niemeyer},
  \citenamefont {Gogolin},\ and\ \citenamefont {Gemmer}}]{thermal1}%
  \BibitemOpen
  \bibfield  {author} {\bibinfo {author} {\bibfnamefont {R.}~\bibnamefont
  {Steinigeweg}}, \bibinfo {author} {\bibfnamefont {A.}~\bibnamefont {Khodja}},
  \bibinfo {author} {\bibfnamefont {H.}~\bibnamefont {Niemeyer}}, \bibinfo
  {author} {\bibfnamefont {C.}~\bibnamefont {Gogolin}}, \ and\ \bibinfo
  {author} {\bibfnamefont {J.}~\bibnamefont {Gemmer}},\ }\href {\doibase
  10.1103/PhysRevLett.112.130403} {\bibfield  {journal} {\bibinfo  {journal}
  {Phys. Rev. Lett.}\ }\textbf {\bibinfo {volume} {112}},\ \bibinfo {pages}
  {130403} (\bibinfo {year} {2014})}\BibitemShut {NoStop}%
\bibitem [{\citenamefont {Goldstein}\ \emph {et~al.}(2006)\citenamefont
  {Goldstein}, \citenamefont {Lebowitz}, \citenamefont {Tumulka},\ and\
  \citenamefont {Zangh\`{\i}}}]{thermal2}%
  \BibitemOpen
  \bibfield  {author} {\bibinfo {author} {\bibfnamefont {S.}~\bibnamefont
  {Goldstein}}, \bibinfo {author} {\bibfnamefont {J.~L.}\ \bibnamefont
  {Lebowitz}}, \bibinfo {author} {\bibfnamefont {R.}~\bibnamefont {Tumulka}}, \
  and\ \bibinfo {author} {\bibfnamefont {N.}~\bibnamefont {Zangh\`{\i}}},\
  }\href {\doibase 10.1103/PhysRevLett.96.050403} {\bibfield  {journal}
  {\bibinfo  {journal} {Phys. Rev. Lett.}\ }\textbf {\bibinfo {volume} {96}},\
  \bibinfo {pages} {050403} (\bibinfo {year} {2006})}\BibitemShut {NoStop}%
\bibitem [{\citenamefont {Landauer}(1961)}]{Landauer}%
  \BibitemOpen
  \bibfield  {author} {\bibinfo {author} {\bibfnamefont {R.}~\bibnamefont
  {Landauer}},\ }\href@noop {} {\bibfield  {journal} {\bibinfo  {journal} {IBM
  Journal of Research and Development}\ }\textbf {\bibinfo {volume} {5}},\
  \bibinfo {pages} {183} (\bibinfo {year} {1961})}\BibitemShut {NoStop}%
\bibitem [{\citenamefont {Bennett}(2003)}]{Bennett_Land}%
  \BibitemOpen
  \bibfield  {author} {\bibinfo {author} {\bibfnamefont {C.~H.}\ \bibnamefont
  {Bennett}},\ }\href {\doibase 10.1016/s1355-2198(03)00039-x} {\bibfield
  {journal} {\bibinfo  {journal} {Studies in History and Philosophy of Modern
  Physics}\ }\textbf {\bibinfo {volume} {34B}},\ \bibinfo {pages} {501}
  (\bibinfo {year} {2003})}\BibitemShut {NoStop}%
\bibitem [{\citenamefont {Bell}(1964)}]{Bell1}%
  \BibitemOpen
  \bibfield  {author} {\bibinfo {author} {\bibfnamefont {J.~S.}\ \bibnamefont
  {Bell}},\ }\href {\doibase 10.1103/PhysicsPhysiqueFizika.1.195} {\bibfield
  {journal} {\bibinfo  {journal} {Physics Physique Fizika}\ }\textbf {\bibinfo
  {volume} {1}},\ \bibinfo {pages} {195} (\bibinfo {year} {1964})}\BibitemShut
  {NoStop}%
\bibitem [{\citenamefont {Bell}(1966)}]{Bell}%
  \BibitemOpen
  \bibfield  {author} {\bibinfo {author} {\bibfnamefont {J.~S.}\ \bibnamefont
  {Bell}},\ }\href {\doibase 10.1103/RevModPhys.38.447} {\bibfield  {journal}
  {\bibinfo  {journal} {Reviews of Modern Physics}\ }\textbf {\bibinfo {volume}
  {38}},\ \bibinfo {pages} {447} (\bibinfo {year} {1966})}\BibitemShut
  {NoStop}%
\bibitem [{\citenamefont {Kraus}(1983)}]{Kraus}%
  \BibitemOpen
  \bibfield  {author} {\bibinfo {author} {\bibfnamefont {K.}~\bibnamefont
  {Kraus}},\ }\href@noop {} {\emph {\bibinfo {title} {States, Effects and
  Operations: Fundamental Notions of Quantum Theory}}}\ (\bibinfo  {publisher}
  {Springer Verlag},\ \bibinfo {year} {1983})\BibitemShut {NoStop}%
\bibitem [{\citenamefont {Nielsen}\ and\ \citenamefont
  {Chuang}(2000)}]{NielsenChuang}%
  \BibitemOpen
  \bibfield  {author} {\bibinfo {author} {\bibfnamefont {M.}~\bibnamefont
  {Nielsen}}\ and\ \bibinfo {author} {\bibfnamefont {I.}~\bibnamefont
  {Chuang}},\ }\href@noop {} {\emph {\bibinfo {title} {Quantum Computation and
  Quantum Information}}}\ (\bibinfo  {publisher} {Cambridge University Press},\
  \bibinfo {year} {2000})\BibitemShut {NoStop}%
\bibitem [{\citenamefont {Peres}(1993)}]{Peres}%
  \BibitemOpen
  \bibfield  {author} {\bibinfo {author} {\bibfnamefont {A.}~\bibnamefont
  {Peres}},\ }\href@noop {} {\emph {\bibinfo {title} {Quantum Theory: Concepts
  and Methods}}}\ (\bibinfo  {publisher} {Kluwer Academic Publishers},\
  \bibinfo {year} {1993})\BibitemShut {NoStop}%
\bibitem [{\citenamefont {Kim}\ \emph {et~al.}(2000)\citenamefont {Kim},
  \citenamefont {Yu}, \citenamefont {Kulik}, \citenamefont {Shih},\ and\
  \citenamefont {Scully}}]{eraser2000}%
  \BibitemOpen
  \bibfield  {author} {\bibinfo {author} {\bibfnamefont {Y.~H.}\ \bibnamefont
  {Kim}}, \bibinfo {author} {\bibfnamefont {R.}~\bibnamefont {Yu}}, \bibinfo
  {author} {\bibfnamefont {S.~P.}\ \bibnamefont {Kulik}}, \bibinfo {author}
  {\bibfnamefont {Y.}~\bibnamefont {Shih}}, \ and\ \bibinfo {author}
  {\bibfnamefont {M.~O.}\ \bibnamefont {Scully}},\ }\href@noop {} {\bibfield
  {journal} {\bibinfo  {journal} {Physical Review Letters}\ }\textbf {\bibinfo
  {volume} {84}},\ \bibinfo {pages} {1} (\bibinfo {year} {2000})}\BibitemShut
  {NoStop}%
\bibitem [{\citenamefont {Walborn}\ \emph {et~al.}(2002)\citenamefont
  {Walborn}, \citenamefont {Cunha}, \citenamefont {Padua},\ and\ \citenamefont
  {Monken}}]{eraser2002}%
  \BibitemOpen
  \bibfield  {author} {\bibinfo {author} {\bibfnamefont {S.~P.}\ \bibnamefont
  {Walborn}}, \bibinfo {author} {\bibfnamefont {M.~O.~T.}\ \bibnamefont
  {Cunha}}, \bibinfo {author} {\bibfnamefont {S.}~\bibnamefont {Padua}}, \ and\
  \bibinfo {author} {\bibfnamefont {C.~H.}\ \bibnamefont {Monken}},\
  }\href@noop {} {\bibfield  {journal} {\bibinfo  {journal} {Physical Review
  A}\ }\textbf {\bibinfo {volume} {65}},\ \bibinfo {pages} {033818} (\bibinfo
  {year} {2002})}\BibitemShut {NoStop}%
\bibitem [{\citenamefont {Korotkov}\ and\ \citenamefont
  {Jordan}(2006)}]{Weak_undoing2006}%
  \BibitemOpen
  \bibfield  {author} {\bibinfo {author} {\bibfnamefont {A.~N.}\ \bibnamefont
  {Korotkov}}\ and\ \bibinfo {author} {\bibfnamefont {A.~N.}\ \bibnamefont
  {Jordan}},\ }\href@noop {} {\bibfield  {journal} {\bibinfo  {journal}
  {Physical Review Letters}\ }\textbf {\bibinfo {volume} {97}},\ \bibinfo
  {pages} {166805} (\bibinfo {year} {2006})}\BibitemShut {NoStop}%
\bibitem [{\citenamefont {Katz}\ \emph {et~al.}(2008)\citenamefont {Katz},
  \citenamefont {Neeley}, \citenamefont {Ansmann}, \citenamefont {Bialczak},
  \citenamefont {Hofheinz}, \citenamefont {Lucero}, \citenamefont {O'Connell},
  \citenamefont {Wang}, \citenamefont {Cleland}, \citenamefont {Martinis},\
  and\ \citenamefont {Korotkov}}]{Weak_2008}%
  \BibitemOpen
  \bibfield  {author} {\bibinfo {author} {\bibfnamefont {N.}~\bibnamefont
  {Katz}}, \bibinfo {author} {\bibfnamefont {M.}~\bibnamefont {Neeley}},
  \bibinfo {author} {\bibfnamefont {M.}~\bibnamefont {Ansmann}}, \bibinfo
  {author} {\bibfnamefont {R.~C.}\ \bibnamefont {Bialczak}}, \bibinfo {author}
  {\bibfnamefont {M.}~\bibnamefont {Hofheinz}}, \bibinfo {author}
  {\bibfnamefont {E.}~\bibnamefont {Lucero}}, \bibinfo {author} {\bibfnamefont
  {A.}~\bibnamefont {O'Connell}}, \bibinfo {author} {\bibfnamefont
  {H.}~\bibnamefont {Wang}}, \bibinfo {author} {\bibfnamefont {A.~N.}\
  \bibnamefont {Cleland}}, \bibinfo {author} {\bibfnamefont {J.~M.}\
  \bibnamefont {Martinis}}, \ and\ \bibinfo {author} {\bibfnamefont {A.~N.}\
  \bibnamefont {Korotkov}},\ }\href@noop {} {\bibfield  {journal} {\bibinfo
  {journal} {Physical Review Letters}\ }\textbf {\bibinfo {volume} {101}},\
  \bibinfo {pages} {200401} (\bibinfo {year} {2008})}\BibitemShut {NoStop}%
\bibitem [{\citenamefont {Griffiths}(1984)}]{hist_grif}%
  \BibitemOpen
  \bibfield  {author} {\bibinfo {author} {\bibfnamefont {R.~B.}\ \bibnamefont
  {Griffiths}},\ }\href {\doibase 10.1007/bf01015734} {\bibfield  {journal}
  {\bibinfo  {journal} {Journal of Statistical Physics}\ }\textbf {\bibinfo
  {volume} {36}},\ \bibinfo {pages} {219} (\bibinfo {year} {1984})}\BibitemShut
  {NoStop}%
\bibitem [{\citenamefont {Isham}(1994)}]{hist_Isham}%
  \BibitemOpen
  \bibfield  {author} {\bibinfo {author} {\bibfnamefont {C.~J.}\ \bibnamefont
  {Isham}},\ }\href {\doibase 10.1063/1.530544} {\bibfield  {journal} {\bibinfo
   {journal} {Journal of Mathematical Physics}\ }\textbf {\bibinfo {volume}
  {35}},\ \bibinfo {pages} {2157} (\bibinfo {year} {1994})}\BibitemShut
  {NoStop}%
\bibitem [{\citenamefont {Omnes}(1992)}]{hist_omnes}%
  \BibitemOpen
  \bibfield  {author} {\bibinfo {author} {\bibfnamefont {R.}~\bibnamefont
  {Omnes}},\ }\href {\doibase 10.1103/RevModPhys.64.339} {\bibfield  {journal}
  {\bibinfo  {journal} {Reviews of Modern Physics}\ }\textbf {\bibinfo {volume}
  {64}},\ \bibinfo {pages} {339} (\bibinfo {year} {1992})}\BibitemShut
  {NoStop}%
\bibitem [{\citenamefont {Dowker}\ and\ \citenamefont
  {Halliwell}(1992)}]{decoh_hist}%
  \BibitemOpen
  \bibfield  {author} {\bibinfo {author} {\bibfnamefont {H.~F.}\ \bibnamefont
  {Dowker}}\ and\ \bibinfo {author} {\bibfnamefont {J.~J.}\ \bibnamefont
  {Halliwell}},\ }\href {\doibase 10.1103/PhysRevD.46.1580} {\bibfield
  {journal} {\bibinfo  {journal} {Physical Review D}\ }\textbf {\bibinfo
  {volume} {46}},\ \bibinfo {pages} {1580} (\bibinfo {year}
  {1992})}\BibitemShut {NoStop}%
\end{thebibliography}%

\end{document}